\documentclass[runningheads]{llncs}
\usepackage{times}

\usepackage{soul}
\usepackage{url}
\usepackage[hidelinks]{hyperref}
\usepackage[utf8]{inputenc}
\usepackage[small]{caption}
\usepackage[ruled,linesnumbered,noend]{algorithm2e}
\usepackage{graphicx}
\usepackage{amsfonts}
\SetKwComment{Comment}{$\ddagger$\ }{}
\usepackage[table]{xcolor}
\usepackage{amsmath}
\usepackage{booktabs}
\begin{document}
\title{Retrieval-efficiency trade-off of Unsupervised \\Keyword Extraction}
%
%
\author{Bla\v{z} \v{S}krlj\inst{1} \and
Boshko Koloski\inst{1} \and
Senja Pollak\inst{1}}
\authorrunning{\v{S}krlj et al.}
%
\institute{Jo\v{z}ef Stefan Institute, Ljubljana, Slovenia
\email{blaz.skrlj@ijs.si}}
\maketitle              
%
\begin{abstract}
Efficiently identifying keyphrases that represent a given document is a challenging task. In the last years, plethora of keyword detection approaches were proposed. These approaches can be based on statistical (frequency-based) properties of e.g., tokens, specialized neural language models, or a graph-based structure derived from a given document. The graph-based methods can be computationally amongst the most efficient ones, while maintaining the retrieval performance. One of the main properties, common to graph-based methods, is their immediate conversion of token space into graphs, followed by subsequent processing. In this paper, we explore a novel unsupervised approach which merges parts of a document in sequential form, \emph{prior to} construction of the token graph. Further, by leveraging personalized PageRank, which considers frequencies of such sub-phrases alongside token lengths during node ranking, we demonstrate state-of-the-art retrieval capabilities while being up to two orders of magnitude faster than current state-of-the-art unsupervised detectors such as YAKE and MultiPartiteRank. The proposed method's scalability was also demonstrated by computing keyphrases for a biomedical corpus comprised of 14 million documents in less than a minute.\\\\
\textbf{Note: The paper was accepted to DS2022}.

\end{abstract}

\section{Introduction}
With the increasing amounts of freely available text-based data sets, methods for efficient keyphrase detection are becoming of high relevance~\cite{hasan2014automatic}. These methods, given a single or multiple documents, output a ranked list of short phrases (or single tokens), which represents\ key aspects of the input text. In the recent years, plethora of keyphrase extraction methods were presented; broadly, they can be divided into unsupervised and supervised ones. This paper focuses on unsupervised keyphrase extraction, i.e. the process where no training set of document is needed to learn to estimate keyphrases -- they are estimated solely based on statistical/topological properties of a given document. The unsupervised methods can be further divided to the ones which construct a graph based on token co-occurrences and the ones which leverage statistical properties of n-grams~\cite{papagiannopoulou2020review}. Recently, neural language model-based keyphrase extraction was also proposed~\cite{grootendorst2020keybert}.
With the abundance of methods, optimization of a single metric becomes less relevant -- methods which maximize e.g., F1@k are common. This paper aims to inform the reader that a realm of highly relevant properties beyond simple retrieval performance can be meaningful in practice, and should be the focus of any novel method proposed (including the adaptation of an existing one presented in this paper). The contributions of this paper are multifold:
\begin{figure}
    \centering
    \includegraphics[width=.9\linewidth]{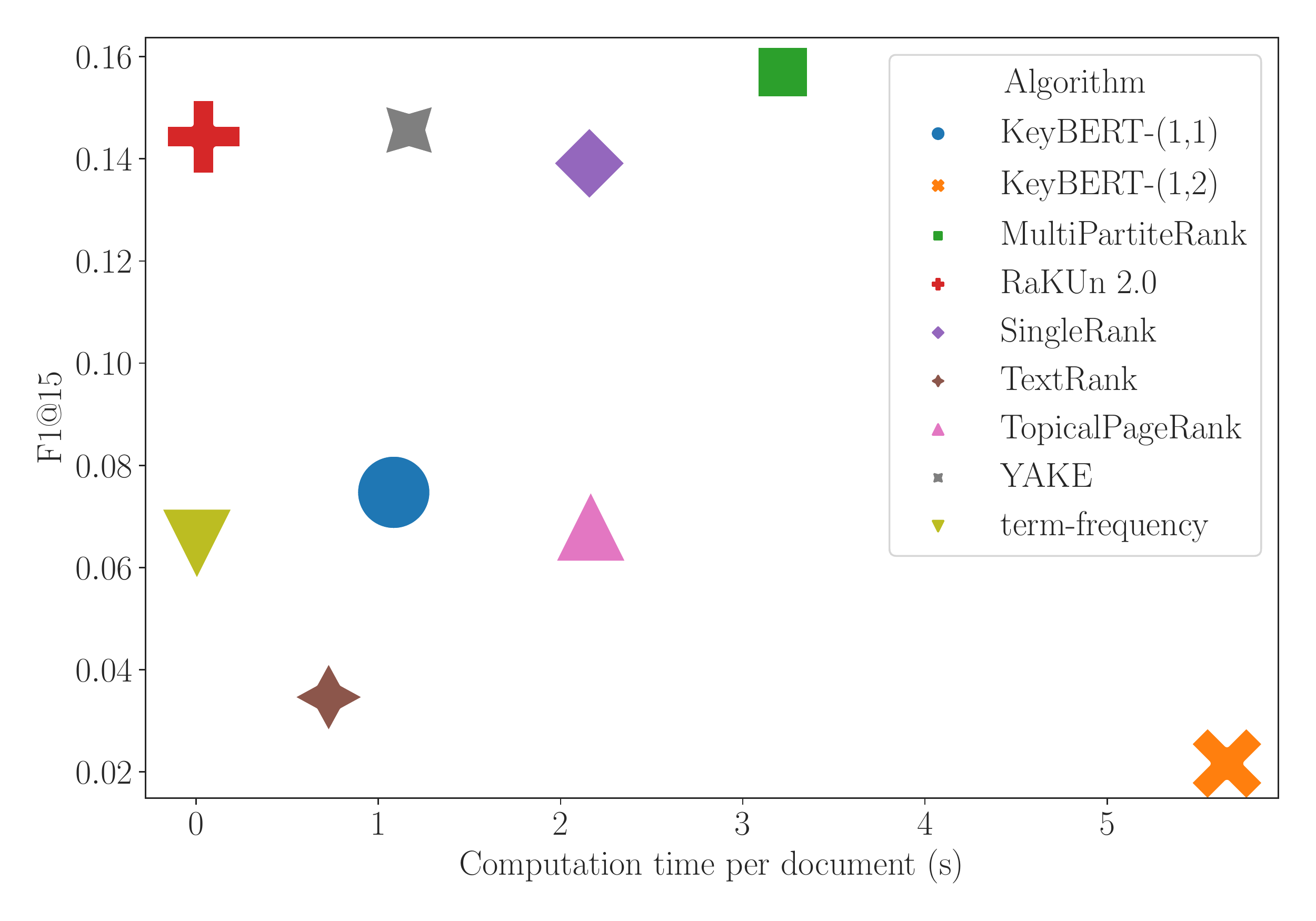}
    \caption{Performance trade-off (time vs. performance) of keyphrase detection methods averaged across fifteen data sets.}
    \label{fig:pareto-overview}
\end{figure}
\begin{enumerate}
    \item We present RaKUn 2.0, a graph-based keyphrase extractor optimized for retrieval-efficiency optimality when considering both retrieval capabilities and performance.
    \item A polygon-based visualization suitable for studying and comparing multiple criteria for multiple keyphrase detection algorithms.
    \item An extensive benchmark of RaKUn 2.0 against strong baselines (including e.g., the recently introduced KeyBERT).
    \item Friedman-Nemenyi-based analysis of average ranks of the algorithms (and their similarity).
\end{enumerate}

\section{Selected related work}
\label{sec:related-work}
\begin{figure}[htb]
    \centering
    \includegraphics[width=.6\linewidth]{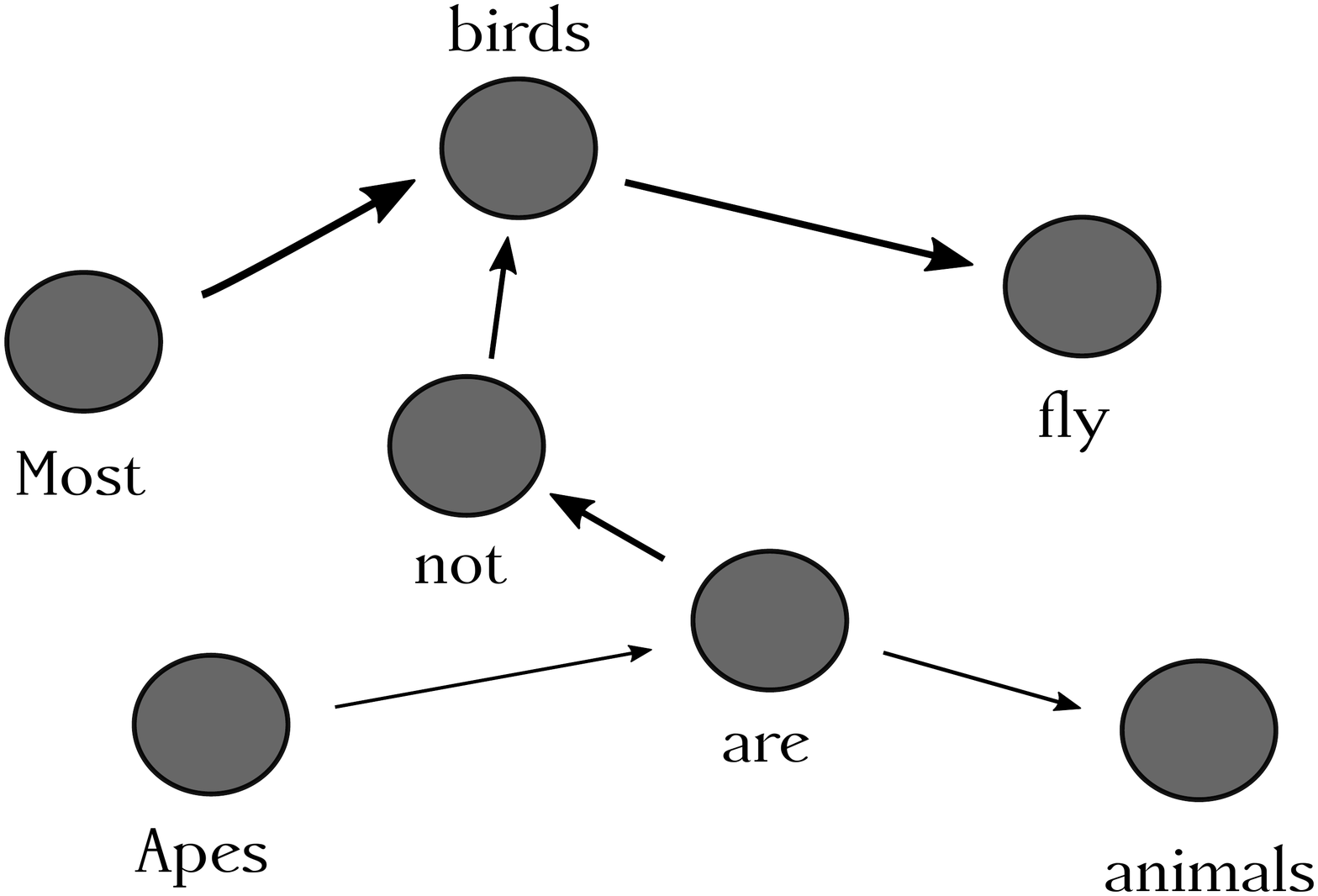}
    \caption{An example token graph.}
    \label{fig:token-graph}
\end{figure}

This section contains an overview of the existing keyphrase detection methods, key underlying ideas and possible caveats of different paradigms. This paper focuses exclusively on \emph{unsupervised} keyphrase extraction -- the process of transforming an input document $D$ in to a ranked collection of keyphrases, i.e. $K=\{(p, s)_k\}; s_{k+1} \leq s_k$, where $k$ represents the top $k$ hits (detected keyphrases), $p$ a given keyphrase and $s$ a given keyphrase's score. The first branch of approaches are based on text-to-graph transformations, followed by subsequent processing of the obtained graphs. Such methods are able to exploit multilevel structure of a document~\cite{boudin2018unsupervised} (MultiPartiteRank), hierarchical structure~\cite{wan-xiao-2008-collabrank} (SingleRank). An example token graph is shown in Figure~\ref{fig:token-graph}.
One of the first graph-based methods was TextRank~\cite{mihalcea-tarau-2004-textrank}, which demonstrated the robustness of graph-based keyphrase detection (and was one of the first to do so). More involved approaches, capable of incorporating topic-level information were also proposed~\cite{bougouin-etal-2013-topicrank} (TopicalPageRank). One of the key issues with graph-based representations is that of node denoising -- the process of identifying the relevant space of nodes which are commonly subject to ranking.
The graph-based methods are highly dependant on the graph construction approach (based on co-occurrence, syntactic, semantic  and similarity information) and node ranking algorithm (e.g. degree, closeness, Page Rank, selectivity, etc.) \cite{overviewKeywordextractionGraphs}. A detailed overview of graph-based methods for keyword extraction and various node-ranking measures is provided in \cite{overviewKeywordextractionGraphs}.

Alongside graph-based methods, statistical methods are also actively developed. One of the most recent examples includes YAKE!~\cite{CAMPOS2020257}, an approach which considers large amounts of n-gram patterns and scores them so that they represent relevant keyphrases. It operates by extracting statistical features from single documents to select the most important keywords of a text. Keyphrase detection was also considered as a task solvable by considering neural language models~\cite{grootendorst2020keybert}. 
An example of this family of models is AttentionRank~\cite{ding-luo-2021-attentionrank}, which exploits the transformer-based neural language model to extract relevant keywords.
A more detailed overview of general keyword detection methods is given in ~\cite{kumar2022comprehensive}.

The discussed approaches seldom focus on metrics beyond retrieval capabilities (e.g., Precision, Recall and F1). One of the purposes of this paper is a comprehensive evaluation of the discussed algorithms with regards to multiple criteria, including computation time and duplication rates (how frequent is a token amongst the space of detected keyphrases).
\section{Proposed algorithm}
\label{sec:rakun2}
The proposed approach sources the core idea from the recent paper on meta vertex-based keyphrase detection RaKUn~\cite{10.1007/978-3-030-31372-2_26}. The considered extension, proposed in this paper is optimized specifically to push the boundary of the retrieval-efficiency front between retrieval performance and retrieval time. We begin with a general overview of the algorithm, followed by theoretical analysis of its complexity (space and time). We refer to the proposed approach as RaKUn 2.0. A high-level overview is shown as Algorithm~\ref{ref:algo-rakun2}.
\begin{algorithm}[ht!]
\DontPrintSemicolon
\caption{RaKUn 2.0}
\label{ref:algo-rakun2}
\KwData{Input document $D$, merge factor $\tau$}
 \textrm{tokens} $\leftarrow$ \textrm{tokenizeDocument}($D$)  \Comment*[r]{Tokenization.}
  \textrm{tokens} $\leftarrow$ \textrm{mergeTokens}($D$)  \Comment*[r]{Merging.}
 $G \leftarrow$ \textrm{documentGraph}(\textrm{tokens})  \Comment*[r]{Weighted graph.}
 $f \leftarrow$ \textrm{tokenFrequencies}(\textrm{tokens})\;
 \textrm{tokenRanks} $\leftarrow$ \textrm{personalizedPR}($G$,$f$)  \Comment*[r]{Ranking.}
 $K \leftarrow$ \textrm{sort}($N(G)$, tokenRanks)  \Comment*[r]{Sorting.}
\KwRet K;
\end{algorithm}
The main steps include \emph{tokenization}, \emph{token merging}, \emph{document graph construction} and \emph{node ranking}. Instead of first constructing (larger) graphs which are subject to node merging into meta vertices, RaKUn 2.0 conducts the merging step at the sequence level, making it more efficient. This step was considered based on an observation that pre-merging tokens in close proximity already offers sufficient results -- by considering only tokens close to one another, no specialized metric for string comparison (possibly expensive) was needed, which substantially sped up the detection process. The second idea which substantially sped up the process is related to \emph{bi-gram hashing}. It refers to constructing a mapping between each bi-gram and its count in the document, enabling fast lookup of this information as follows. for each subsequent token pair ($t_i, t_j$) term counts are retrieved (they are pre-computed during tokenization). We next compute a \emph{merge threshold} score as:
\begin{equation*}
    \textrm{MScore} = \frac{|\#t_i - \#b_{ij}| + |\#t_j - \#b_{ij}|}{\#t_i + \#t_j}\;
\end{equation*}
\noindent where $t_i$ and $t_j$ are two subsequent tokens, and $b_{ij}$ is the bi-gram comprised of the two tokens. If MScore is lower than a user-specified threshold (hyperparameter), the merged token is added as a new token to the token space, and term counts of the two individual tokens are diminished by MScore as $\#t_i = \textrm{MScore} * \#t_i$, i.e., multiplied with the computed score. Values of MScore, lower than one, imply more emphasis of multi-term keyphrases (individual terms are not as emphasized), and values larger than one imply more individual token keyphrases.  Hence, the MScore serves as an intermediary step which \emph{emphasizes} specific tokens during the ranking step.

The token graph $G$ is constructed from the modified list of tokens by considering subsequent, lower-cased tokens as edges. The edge weights are incremented every time a given bi-gram repeats -- the transitions between tokens which commonly co-occur are emphasized. The next step is \emph{node ranking}. Here, a real-valued score is assigned to each (pre-merged) token. We consider personalized PageRank algorithm~\cite{ilprints422}, where the personalization vector is constructed based on term counts. This step results in real-valued scores (between 0 and 1) for each token. The final set of scores is obtained by computing an element-wise product between the PageRank scores and token lengths. This step emphasizes longer keyphrases. We traverse the space of scored tokens and remove case-level duplicates (e.g., `City' and `city').

The described algorithm for keyphrase detection was conceived with simplicity in mind. This property also resonates with its computational complexity. Let $T$ represent the number of tokens after the merge step (cardinality difference is negligible with regards to the runtime). Both graph construction and merging need one pass across the token sequence ($\mathcal{O}(|T|)$. The computationally most expensive part is computation of personalized PageRank. In theory, PageRank's complexity is $\mathcal{O}(|T| + l)$, where $l$ is the number of links in the constructed token graph. In practice, the obtained graphs are very sparse -- only selected bi-grams co-occur. The opposite case, where dense, clique-like graphs would be produced would imply appearance of tokens in highly diverse contexts, which is highly unlikely. 
The final step requires sorting of tokens based their scores. This yields the final complexity of $\mathcal{O}(|T|\log{|T|} \cdot l)$.
Assuming very sparse graphs (as observed during the experiments), the complexity remains linear with regards to the number of tokens in the token set after the merge step.

\section{Evaluation}
We next discuss the evaluation procedures used to estimate the performance of individual algorithms, followed by a discussion regarding their comparison.
We evaluate each algorithm with regards to three main aspects; retrieval performance, keyword duplication rate and computation time. The retrieval performance was measured as done in the previous work~\cite{CAMPOS2020257}. Precision@k is defined as $\frac{|\textrm{Gold} \cap \textrm{k-predicted}|}{k}$. Recall@k is defined as $\frac{|\textrm{Gold} \cap \textrm{k-predicted}|}{|\textrm{Gold}|}$. Precision represents the number of keyphrases retrieved with regards to top $k$ predicted ones, while recall represents the overall retrieval capability. We also computed (macro) F1, which is the harmonic mean of precision and recall, averaged across documents.

The second score is the \emph{duplication rate}. We compute this score as follows;
for each detected keyphrase, we first split it to separate tokens (if multi-token keyphrase is considered). For each part, we traverse the space of detected tokens.
If there is a match, we increment a duplicate counter, otherwise, we increment the non\_duplicate counter. The final score is computed as $\frac{\#duplicates}{\#non\_duplicates + 1}$, and was observed to be in the interval [0, 1]. The computation time was measured in seconds (for each document).
\begin{table}[h!]
    \centering
    \caption{Summary of the considered data sets.}
    \resizebox{0.5\textwidth}{!}{
\begin{tabular}{lrrrr}
\toprule
          Dataset &  \#Docs &  \#KW &  Mean KW tokens &  Mean doc len \\
\midrule
           wiki20~\cite{medelyan2008topic} &     20 & 35.5 &             2.0 &        7728.0 \\
            fao30~\cite{https://doi.org/10.1002/asi.20790} &     30 & 32.2 &             1.6 &        4710.3 \\
        theses100~\cite{medelyan2009human} &    100 &  6.7 &             2.0 &        4813.9 \\
     citeulike180~\cite{10.5555/1699648.1699678} &    183 & 17.4 &             1.3 &        4517.9 \\
       Nguyen2007~\cite{10.1007/978-3-540-77094-7_41} &    209 & 12.0 &             2.1 &        4425.6 \\
      SemEval2010~\cite{10.5555/1859664.1859668} &    243 & 15.6 &             2.2 &        7093.3 \\
      SemEval2017~\cite{augenstein2017semeval} &    493 & 17.3 &             2.9 &         168.3 \\
500N-KPCrowd-v1.1~\cite{marujo2013keyphrase} &    500 & 49.2 &             1.4 &         393.9 \\
           PubMed~\cite{aronson2000nlm} &    500 & 14.2 &             1.9 &        3880.2 \\
              kdd~\cite{gollapalli2014extracting} &    755 &  4.1 &             2.0 &          74.1 \\
           fao780~\cite{https://doi.org/10.1002/asi.20790} &    779 &  8.0 &             1.6 &        4685.0 \\
       Schutz2008~\cite{schutz2008keyphrase} &   1231 & 45.3 &             1.5 &        2362.6 \\
              www~\cite{gollapalli2014extracting} &   1330 &  4.8 &             1.9 &          82.0 \\
           Inspec~\cite{10.3115/1119355.1119383} &   2000 & 14.1 &             2.2 &         112.5 \\
     Krapivin2009~\cite{krapivin2009large} &   2304 &  5.3 &             2.1 &        7094.1 \\
\bottomrule
\end{tabular}
}
    \label{tab:datasets-summary}
\end{table}
For visualization of retrieval-efficiency tradeoffs with regards to the mentioned scores it makes sense to have uniform meaning of large and small values. Hence, we introduce the following adapted scores which reflect this idea. The retrieval capability already corresponds to e.g., F1 score, meaning that higher values are preferred. We additionally normalize F1 scores to range between 0 and 1 based on the worst-best performing algorithms (on average). This way, an algorithm scored with 0 is the worst-performing one, while the top performing is scored with 1 (see Figure 8). 
Similar adaptations were considered for time performance (normalized inverse times) and duplication rates (normalized inverse duplication rates). One of the main results of this paper is a visualization which jointly considers all three aspects.
The considered collection of data sets is summarized in Table~\ref{tab:datasets-summary}.

The considered baselines are discussed next. The graph-based baselines include MultiPartiteRank~\cite{boudin2018unsupervised}, SingleRank~\cite{wan-xiao-2008-collabrank}, TextRank~\cite{mihalcea-tarau-2004-textrank} and TopicalPageRank~\cite{bougouin-etal-2013-topicrank}. The statistical baseline considered was YAKE~\cite{CAMPOS2020257}. The language model-based baseline is the recent KeyBERT~\cite{grootendorst2020keybert}. For all approaches, we considered the default hyperparameter configurations, as we were interested in out-of-the-box performance. We computed, however, two variants of KeyBERT, one which emits single tokens (KeyBERT-(1,1)) and one which permits two term tokens (KeyBERT-(1,2)). Default configuration of KeyBERT variants performed worse than term frequency-based extraction~\footnote{We considered unigrams. Inverse document frequencies were not computed as they require the whole corpus, making them not directly comparable to purely unsupervised methods.}, and offered (1,1) adequate performance only when we set the `maxsum` and `mmr` flags to `true`. The stopwords used were the same for all approaches (NLTK's default English stopwords~\cite{bird2009natural}). Other algorithms' implementations were based on the PKE library~\cite{boudin:2016:COLINGDEMO}. 

\section{Results}
\label{sec:results}
\begin{figure}[t!]
    \centering
    \includegraphics[width=\linewidth]{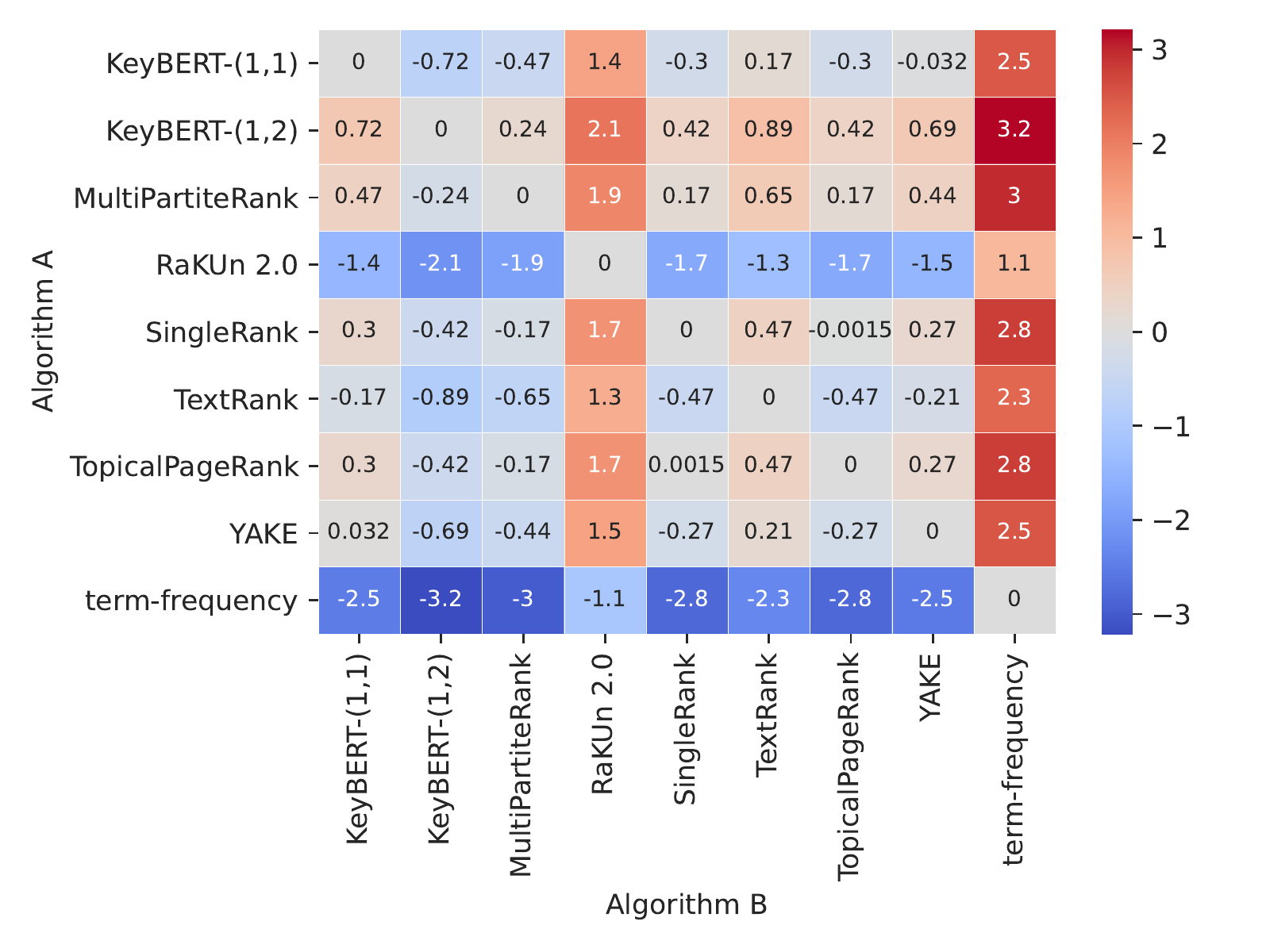}
    \caption{Pairwise time comparison of average algorithm run times ($\log_{10}(\frac{A}{B})$.}
    \label{fig:runtimes-heatmap}
\end{figure}

\begin{figure}[htb]
    \centering
    \includegraphics[width=.85\linewidth]{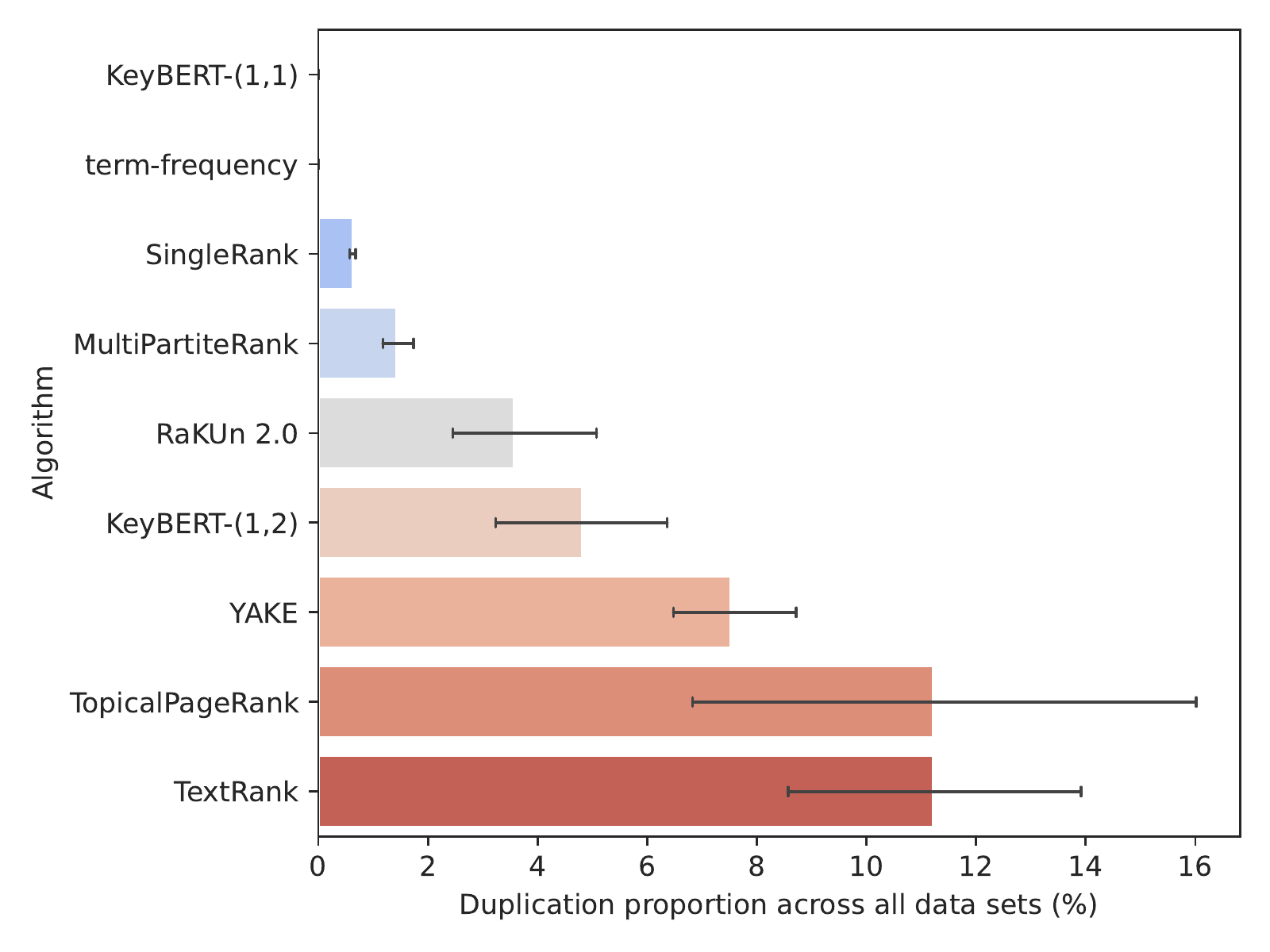}
    \caption{Duplication levels for different algorithms.}
    \label{fig:duplication}
\end{figure}

A summary of algorithm run times (relative to one another) is shown in Figure~\ref{fig:runtimes-heatmap}.
As expected, the simplest baseline (term frequency) is up to three orders of magnitude faster than e.g., BERT-based model. The second approach that performs substantially better, while remaining up to two orders of magnitude faster is the proposed RaKUn 2.0. It is closely followed by SingleRank and TopicalPageRank.
The duplication levels are shown in Figure~\ref{fig:duplication}.
The duplication ablation indicates the highest duplication levels were observed for YAKE, TopicalPageRank and TextRank. MultiPartiteRank and SingleRank had notably lower duplication levels (KeyBERT-(1,1) as well the term frequency (unigram) baseline.
The proposed RaKUn 2.0 is at the lower end of the approaches with regards to this score, albeit not being optimal. 

\begin{figure}[htb]
    \centering
    \includegraphics[width=\linewidth]{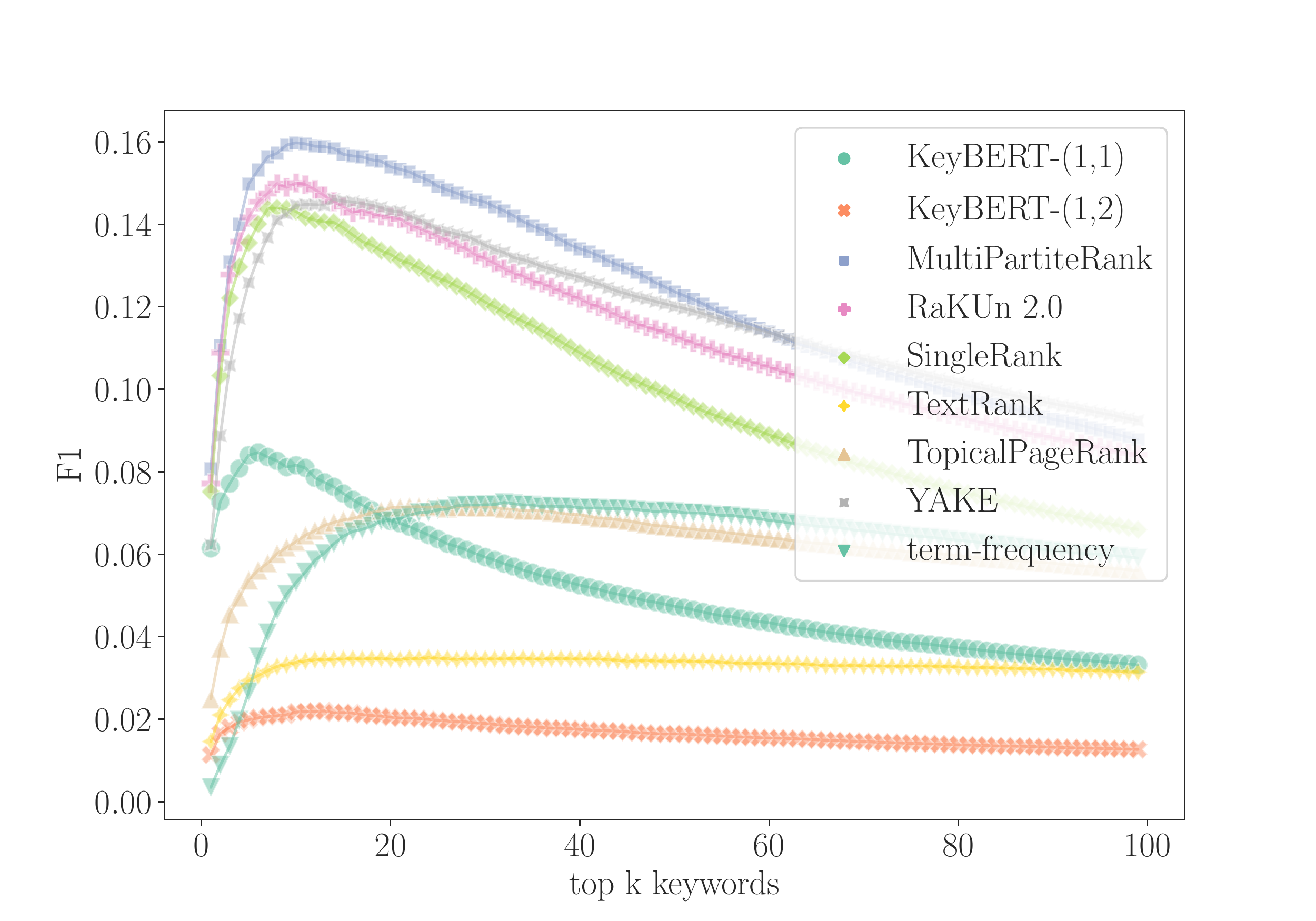}
    \caption{F1 score for different top $k$ keyphrases, averaged across all data sets.}
    \label{fig:overall-f1}
\end{figure}

We continue the discussion by presenting the retrieval performance. A systematic investigation of algorithm performance is shown in Figure~\ref{fig:overall-f1}.
The results indicate that on average, MultiPartiteRank is the leading algorithm in the low $k$ scenarios. RaKUn 2.0, however, performs very similarly for up to ten keyphrases, which is one of the most common usecases of such algorithms. A more detailed overview of the scores on the per-data set level is given in Tables 1-5. The color codes represent top three performers for each data set (gold=first, silver=second, bronze=third).
\begin{table*}[]
    \centering
    \caption{F1@10 (gold=first, silver=second, bronze=third, per row)}
    \resizebox{0.85\textwidth}{!}{
    \begin{tabular}{llllllllll}
\toprule
Algorithm &             KeyBERT-(1,1) & KeyBERT-(1,2) &            MultiPartiteRank &                   RaKUn 2.0 &                 SingleRank & TextRank &             TopicalPageRank &                        YAKE & TFreq \\
Dataset           &                           &               &                             &                             &                            &          &                             &                             &                \\
\midrule
500N-KPCrowd-v1.1 &                     0.012 &         0.012 &  \cellcolor{yellow!30}0.171 &   \cellcolor{brown!30}0.138 &   \cellcolor{gray!30}0.164 &    0.057 &                       0.094 &                       0.127 &          0.106 \\
Inspec            &                       0.0 &           0.0 &     \cellcolor{gray!30}0.22 &                       0.143 &  \cellcolor{brown!30}0.207 &    0.126 &   \cellcolor{yellow!30}0.24 &                       0.195 &          0.041 \\
Krapivin2009      &                     0.051 &         0.057 &    \cellcolor{gray!30}0.109 &   \cellcolor{brown!30}0.097 &                      0.094 &    0.007 &                        0.02 &  \cellcolor{yellow!30}0.118 &          0.011 \\
Nguyen2007        &                     0.099 &         0.058 &    \cellcolor{gray!30}0.168 &                       0.141 &  \cellcolor{brown!30}0.152 &    0.025 &                       0.053 &  \cellcolor{yellow!30}0.188 &          0.035 \\
PubMed            &  \cellcolor{yellow!30}0.1 &         0.021 &   \cellcolor{brown!30}0.087 &                       0.083 &                      0.072 &    0.002 &                       0.004 &    \cellcolor{gray!30}0.087 &          0.036 \\
Schutz2008        &                     0.088 &         0.023 &   \cellcolor{yellow!30}0.23 &   \cellcolor{brown!30}0.194 &   \cellcolor{gray!30}0.219 &    0.015 &                       0.031 &                        0.15 &          0.075 \\
SemEval2010       &                     0.071 &         0.053 &    \cellcolor{gray!30}0.152 &   \cellcolor{brown!30}0.139 &                      0.133 &     0.01 &                       0.023 &  \cellcolor{yellow!30}0.155 &          0.023 \\
SemEval2017       &                       0.0 &           0.0 &    \cellcolor{gray!30}0.216 &                       0.132 &  \cellcolor{brown!30}0.203 &    0.122 &  \cellcolor{yellow!30}0.224 &                       0.175 &          0.056 \\
citeulike180      &  \cellcolor{gray!30}0.205 &          0.03 &                       0.172 &  \cellcolor{yellow!30}0.225 &                       0.14 &    0.004 &                       0.013 &   \cellcolor{brown!30}0.185 &          0.097 \\
fao30             &                      0.16 &         0.027 &    \cellcolor{gray!30}0.176 &  \cellcolor{yellow!30}0.233 &  \cellcolor{brown!30}0.161 &    0.008 &                       0.011 &                        0.15 &          0.072 \\
fao780            &                     0.116 &         0.013 &  \cellcolor{yellow!30}0.141 &   \cellcolor{brown!30}0.138 &                      0.118 &    0.004 &                       0.009 &    \cellcolor{gray!30}0.138 &          0.064 \\
kdd               &                       0.0 &         0.001 &                       0.107 &    \cellcolor{gray!30}0.144 &                      0.094 &    0.058 &   \cellcolor{brown!30}0.109 &  \cellcolor{yellow!30}0.144 &          0.056 \\
theses100         &                     0.099 &         0.017 &  \cellcolor{yellow!30}0.149 &   \cellcolor{brown!30}0.103 &   \cellcolor{gray!30}0.128 &    0.004 &                       0.006 &                       0.093 &          0.042 \\
wiki20            &  \cellcolor{gray!30}0.222 &         0.013 &   \cellcolor{brown!30}0.186 &  \cellcolor{yellow!30}0.226 &                      0.163 &      0.0 &                         0.0 &                       0.135 &          0.021 \\
www               &                       0.0 &         0.001 &    \cellcolor{brown!30}0.11 &    \cellcolor{gray!30}0.113 &                      0.099 &    0.065 &                       0.109 &  \cellcolor{yellow!30}0.129 &          0.062 \\
\bottomrule
\end{tabular}

    }
    \caption{Precision@10 (gold=first, silver=second, bronze=third, per row)}
        \resizebox{0.85\textwidth}{!}{
        \begin{tabular}{llllllllll}
\toprule
Algorithm &               KeyBERT-(1,1) & KeyBERT-(1,2) &            MultiPartiteRank &                   RaKUn 2.0 &                 SingleRank & TextRank &             TopicalPageRank &                        YAKE & TFreq \\
Dataset           &                             &               &                             &                             &                            &          &                             &                             &                \\
\midrule
500N-KPCrowd-v1.1 &                       0.046 &         0.037 &   \cellcolor{yellow!30}0.38 &   \cellcolor{brown!30}0.323 &    \cellcolor{gray!30}0.36 &    0.129 &                       0.173 &                       0.262 &          0.192 \\
Inspec            &                         0.0 &           0.0 &    \cellcolor{gray!30}0.174 &                       0.112 &  \cellcolor{brown!30}0.165 &    0.101 &  \cellcolor{yellow!30}0.189 &                       0.152 &          0.032 \\
Krapivin2009      &                       0.037 &          0.04 &    \cellcolor{gray!30}0.079 &   \cellcolor{brown!30}0.069 &                      0.068 &    0.005 &                       0.014 &  \cellcolor{yellow!30}0.084 &          0.008 \\
Nguyen2007        &                        0.09 &          0.05 &    \cellcolor{gray!30}0.151 &                       0.124 &  \cellcolor{brown!30}0.138 &    0.022 &                       0.048 &  \cellcolor{yellow!30}0.166 &          0.032 \\
PubMed            &  \cellcolor{yellow!30}0.065 &         0.013 &   \cellcolor{brown!30}0.057 &                       0.055 &                      0.047 &    0.001 &                       0.003 &    \cellcolor{gray!30}0.057 &          0.023 \\
Schutz2008        &                       0.193 &         0.047 &  \cellcolor{yellow!30}0.504 &   \cellcolor{brown!30}0.433 &    \cellcolor{gray!30}0.48 &    0.029 &                       0.065 &                       0.329 &          0.163 \\
SemEval2010       &                       0.075 &         0.055 &    \cellcolor{gray!30}0.159 &   \cellcolor{brown!30}0.146 &                       0.14 &    0.011 &                       0.024 &  \cellcolor{yellow!30}0.162 &          0.023 \\
SemEval2017       &                         0.0 &         0.001 &    \cellcolor{gray!30}0.293 &                       0.184 &  \cellcolor{brown!30}0.278 &    0.169 &    \cellcolor{yellow!30}0.3 &                       0.235 &          0.077 \\
citeulike180      &    \cellcolor{gray!30}0.208 &          0.03 &                       0.172 &  \cellcolor{yellow!30}0.228 &                       0.14 &    0.003 &                       0.012 &   \cellcolor{brown!30}0.183 &          0.097 \\
fao30             &                       0.183 &         0.033 &     \cellcolor{gray!30}0.21 &   \cellcolor{yellow!30}0.28 &   \cellcolor{brown!30}0.19 &     0.01 &                       0.013 &                        0.18 &          0.087 \\
fao780            &                       0.075 &         0.008 &  \cellcolor{yellow!30}0.092 &     \cellcolor{gray!30}0.09 &                      0.077 &    0.002 &                       0.006 &   \cellcolor{brown!30}0.089 &          0.041 \\
kdd               &                         0.0 &         0.001 &                       0.064 &  \cellcolor{yellow!30}0.087 &                      0.056 &    0.036 &   \cellcolor{brown!30}0.065 &    \cellcolor{gray!30}0.085 &          0.034 \\
theses100         &                       0.064 &         0.011 &  \cellcolor{yellow!30}0.098 &   \cellcolor{brown!30}0.068 &   \cellcolor{gray!30}0.084 &    0.002 &                       0.004 &                        0.06 &          0.027 \\
wiki20            &     \cellcolor{gray!30}0.19 &          0.01 &   \cellcolor{brown!30}0.155 &   \cellcolor{yellow!30}0.19 &                      0.135 &      0.0 &                         0.0 &                        0.12 &           0.02 \\
www               &                         0.0 &         0.001 &   \cellcolor{brown!30}0.066 &    \cellcolor{gray!30}0.068 &                       0.06 &     0.04 &                       0.065 &  \cellcolor{yellow!30}0.076 &          0.037 \\
\bottomrule
\end{tabular}

        }
    \caption{Recall@10 (gold=first, silver=second, bronze=third, per row)}
         \resizebox{0.85\textwidth}{!}{
    \begin{tabular}{llllllllll}
\toprule
Algorithm &              KeyBERT-(1,1) & KeyBERT-(1,2) &            MultiPartiteRank &                   RaKUn 2.0 &                 SingleRank & TextRank &             TopicalPageRank &                        YAKE & TFreq \\
Dataset           &                            &               &                             &                             &                            &          &                             &                             &                \\
\midrule
500N-KPCrowd-v1.1 &                      0.007 &         0.007 &  \cellcolor{yellow!30}0.144 &                       0.119 &   \cellcolor{gray!30}0.139 &    0.041 &                       0.087 &   \cellcolor{brown!30}0.129 &          0.113 \\
Inspec            &                        0.0 &           0.0 &    \cellcolor{gray!30}0.356 &                       0.233 &  \cellcolor{brown!30}0.331 &    0.194 &  \cellcolor{yellow!30}0.388 &                       0.326 &           0.07 \\
Krapivin2009      &                      0.097 &         0.119 &    \cellcolor{gray!30}0.212 &   \cellcolor{brown!30}0.187 &                      0.182 &    0.014 &                       0.041 &  \cellcolor{yellow!30}0.236 &          0.021 \\
Nguyen2007        &                      0.135 &         0.087 &     \cellcolor{gray!30}0.23 &   \cellcolor{brown!30}0.216 &                      0.205 &    0.036 &                       0.078 &  \cellcolor{yellow!30}0.279 &           0.05 \\
PubMed            &  \cellcolor{yellow!30}0.27 &         0.065 &   \cellcolor{brown!30}0.223 &                       0.209 &                      0.181 &    0.009 &                       0.013 &    \cellcolor{gray!30}0.239 &          0.096 \\
Schutz2008        &                      0.063 &         0.017 &  \cellcolor{yellow!30}0.161 &   \cellcolor{brown!30}0.133 &   \cellcolor{gray!30}0.153 &    0.011 &                       0.022 &                       0.104 &          0.053 \\
SemEval2010       &                      0.071 &         0.054 &    \cellcolor{gray!30}0.152 &   \cellcolor{brown!30}0.139 &                      0.133 &     0.01 &                       0.024 &  \cellcolor{yellow!30}0.156 &          0.023 \\
SemEval2017       &                        0.0 &           0.0 &    \cellcolor{gray!30}0.183 &                        0.11 &   \cellcolor{brown!30}0.17 &    0.101 &  \cellcolor{yellow!30}0.189 &                        0.15 &          0.046 \\
citeulike180      &   \cellcolor{gray!30}0.221 &         0.034 &                       0.187 &  \cellcolor{yellow!30}0.242 &                      0.151 &    0.005 &                       0.016 &   \cellcolor{brown!30}0.205 &          0.104 \\
fao30             &  \cellcolor{brown!30}0.149 &         0.023 &    \cellcolor{gray!30}0.159 &   \cellcolor{yellow!30}0.21 &                      0.147 &    0.007 &                        0.01 &                       0.134 &          0.065 \\
fao780            &                      0.335 &         0.036 &  \cellcolor{yellow!30}0.396 &    \cellcolor{brown!30}0.39 &                      0.321 &     0.01 &                       0.025 &     \cellcolor{gray!30}0.39 &          0.193 \\
kdd               &                        0.0 &         0.003 &                       0.384 &    \cellcolor{gray!30}0.514 &                      0.346 &    0.188 &   \cellcolor{brown!30}0.398 &  \cellcolor{yellow!30}0.562 &          0.194 \\
theses100         &  \cellcolor{brown!30}0.266 &         0.045 &  \cellcolor{yellow!30}0.389 &                       0.262 &   \cellcolor{gray!30}0.326 &    0.019 &                        0.02 &                       0.254 &          0.116 \\
wiki20            &   \cellcolor{gray!30}0.294 &         0.017 &   \cellcolor{brown!30}0.251 &  \cellcolor{yellow!30}0.297 &                      0.221 &      0.0 &                         0.0 &                       0.166 &          0.023 \\
www               &                        0.0 &         0.004 &                       0.393 &    \cellcolor{gray!30}0.412 &                      0.352 &    0.221 &   \cellcolor{brown!30}0.394 &  \cellcolor{yellow!30}0.502 &          0.237 \\
\bottomrule
\end{tabular}

    }
      \caption{Retrieval time (s). (gold=first, silver=second, bronze=third, per row)}
         \resizebox{0.85\textwidth}{!}{
    \begin{tabular}{llllllllll}
\toprule
Algorithm &              KeyBERT-(1,1) & KeyBERT-(1,2) & MultiPartiteRank &                 RaKUn 2.0 & SingleRank &                   TextRank & TopicalPageRank &   YAKE &              TFreq \\
Dataset           &                            &               &                  &                           &            &                            &                 &        &                             \\
\midrule
500N-KPCrowd-v1.1 &                      0.422 &         0.763 &            0.477 &  \cellcolor{gray!30}0.009 &      0.454 &  \cellcolor{brown!30}0.417 &           1.552 &  0.699 &    \cellcolor{yellow!30}0.0 \\
Inspec            &  \cellcolor{brown!30}0.202 &         0.337 &            0.399 &  \cellcolor{gray!30}0.006 &        0.4 &                      0.394 &           1.554 &   0.74 &    \cellcolor{yellow!30}0.0 \\
Krapivin2009      &                      1.561 &        11.282 &            6.144 &   \cellcolor{gray!30}0.08 &      4.332 &  \cellcolor{brown!30}1.111 &           2.852 &  1.423 &  \cellcolor{yellow!30}0.007 \\
Nguyen2007        &                      1.333 &         6.318 &            3.528 &   \cellcolor{gray!30}0.05 &       2.45 &   \cellcolor{brown!30}0.87 &           2.505 &  1.304 &  \cellcolor{yellow!30}0.005 \\
PubMed            &                      1.237 &         4.865 &            2.823 &  \cellcolor{gray!30}0.046 &      1.945 &  \cellcolor{brown!30}0.766 &           2.312 &  1.249 &  \cellcolor{yellow!30}0.004 \\
Schutz2008        &                       1.58 &         6.926 &            3.993 &  \cellcolor{gray!30}0.038 &      2.675 &  \cellcolor{brown!30}0.772 &           2.428 &  1.236 &  \cellcolor{yellow!30}0.004 \\
SemEval2010       &                      1.675 &        12.479 &            6.135 &  \cellcolor{gray!30}0.076 &      4.213 &  \cellcolor{brown!30}1.117 &           2.707 &  1.378 &  \cellcolor{yellow!30}0.008 \\
SemEval2017       &  \cellcolor{brown!30}0.213 &         0.431 &            0.403 &  \cellcolor{gray!30}0.007 &      0.395 &                      0.389 &           1.596 &  0.947 &    \cellcolor{yellow!30}0.0 \\
citeulike180      &                      1.556 &         8.006 &            3.937 &  \cellcolor{gray!30}0.051 &      2.411 &  \cellcolor{brown!30}0.812 &           2.446 &  1.322 &  \cellcolor{yellow!30}0.005 \\
fao30             &                      1.528 &         7.599 &            4.665 &  \cellcolor{gray!30}0.056 &      2.793 &  \cellcolor{brown!30}0.877 &            2.47 &  1.573 &  \cellcolor{yellow!30}0.005 \\
fao780            &                      1.531 &         7.806 &            5.284 &  \cellcolor{gray!30}0.056 &      3.111 &  \cellcolor{brown!30}0.838 &            2.53 &  1.479 &  \cellcolor{yellow!30}0.005 \\
kdd               &  \cellcolor{brown!30}0.153 &         0.268 &            0.394 &  \cellcolor{gray!30}0.006 &      0.394 &                      0.383 &           1.371 &  0.549 &    \cellcolor{yellow!30}0.0 \\
theses100         &                       1.52 &         7.069 &             4.05 &  \cellcolor{gray!30}0.053 &      2.603 &  \cellcolor{brown!30}0.811 &           2.293 &  1.644 &  \cellcolor{yellow!30}0.004 \\
wiki20            &                      1.598 &        10.453 &            5.674 &  \cellcolor{gray!30}0.066 &        3.8 &  \cellcolor{brown!30}0.952 &           2.586 &  1.429 &  \cellcolor{yellow!30}0.006 \\
www               &  \cellcolor{brown!30}0.152 &         0.269 &             0.39 &  \cellcolor{gray!30}0.006 &      0.396 &                      0.396 &           1.283 &  0.552 &    \cellcolor{yellow!30}0.0 \\
\bottomrule
\end{tabular}

    }
    \label{tab:summary-f1}
\end{table*}
We additionally conducted rank-based difference significance evaluation~\cite{JMLR:v7:demsar06a}, where the average algorithm ranks are compared across all data sets. If the algorithms are linked with a red line, they perform very similarly ($p<0.05$). The diagrams are shown as Figures~\ref{fig:cd-f1} and ~\ref{fig:cd-time}.
\begin{figure}[htb]
    \centering
    \includegraphics[width=\linewidth]{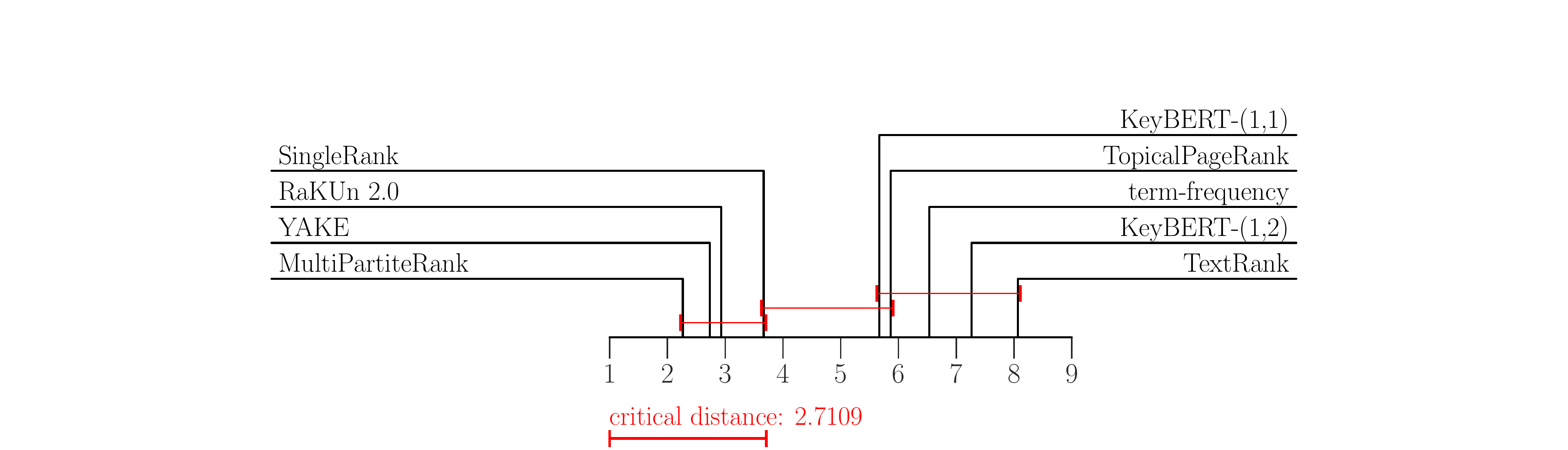}
    \caption{Critical difference diagram - F1@15. RaKUn 2.0's performance is (statistically) comparable to the recent state-of-the-art approaches.}
        \label{fig:cd-f1}
        \includegraphics[width=\linewidth]{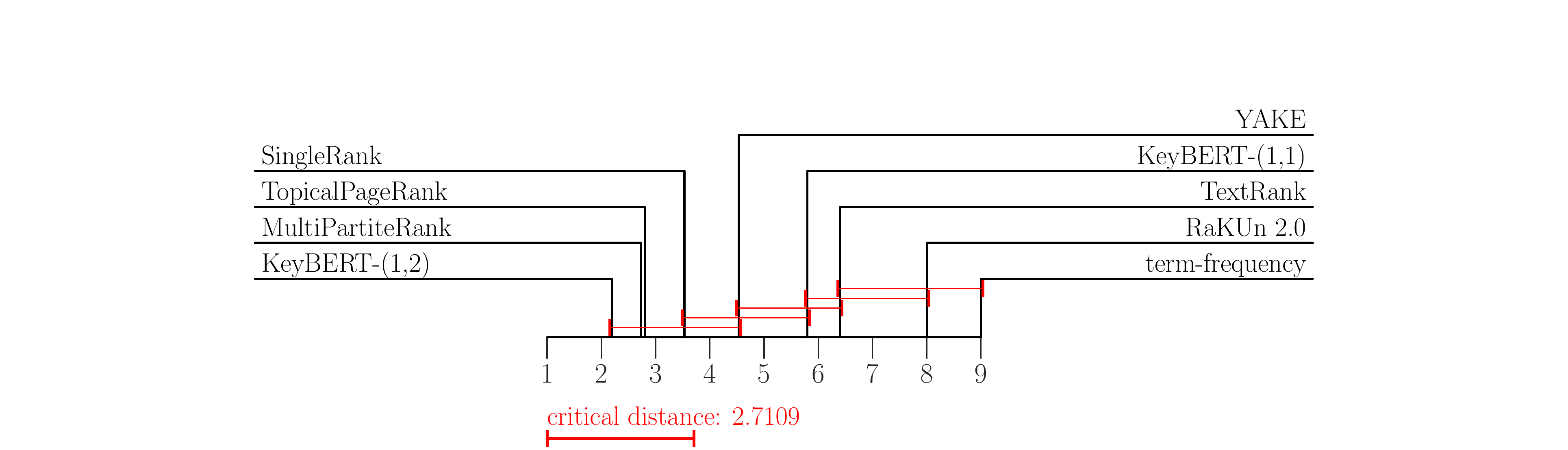}
    \caption{CD diagrams -- time per document. Higher ranks indicate faster compute time. RaKUn 2.0 is significantly faster when compared to other state-of-the-art methods.}
    \label{fig:cd-time}
\end{figure}
The tests indicate that the difference between the top-performing approaches (MultiPartiteRank, YAKE and RaKUn 2.0) is insignificant. Similar observations can be made based on tabular summaries. Overall, however, we can observe a marginal dominance of RaKUn 2.0 w.r.t. precision. Similar retrieval performance amplifies the purpose of this paper, which transcends the retrieval-only evaluation and incorporates also other properties of either the algorithms or the retrieved space.

\begin{figure}[htb!]
    \centering
    \includegraphics[width=\linewidth]{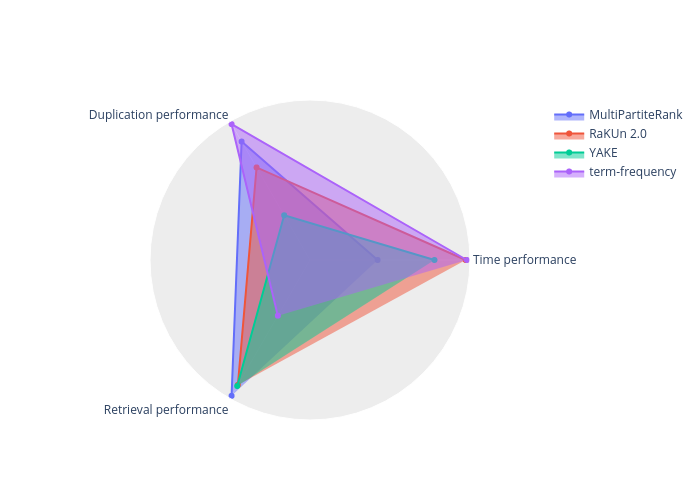}
    \caption{A visualization comparing best and worst-performing approaches with regards to three different criteria relevant in practice. Note that the scores are relative with regards to the considered methods' performances.}
    \label{fig:pareto-poly}
\end{figure}

In Figure~\ref{fig:pareto-poly}, the selected approaches are compared across the three main evaluation criteria -- retrieval performance, duplication performance (inverse of duplication rate) and time performance (inverse of normalized times across all algorithms). Larger values are better for each criterion. It can be observed that MultiPartiteRank outperforms the others at the front considering duplication and retrieval performance, however, RaKUn 2.0 outperforms the others when considering retrieval capabilities and computation time.

\subsection{Scaling to 14M documents}
A direct way of testing the complexity bounds stated in the methods section was to attempt and run RaKUn 2.0 directly on the collection of approximately 14 million biomedical articles -- the MeDAL corpus~\cite{wen-etal-2020-medal}\footnote{\url{https://www.reddit.com/r/MachineLearning/comments/jx63fd/r_a_14m_articles_dataset_for_medical_nlp/}}. The corpus was parsed into a list of documents and fed into the default configuration of RaKUn 2.0. The computation took approximately forty seconds (including text reading) on a virtual machine with 12 cores and 32GB of RAM. The list of top ten keyphrases is shown as Table~\ref{tab:top-medal}.

\begin{table}[htb!]
    \centering
       \caption{14M articles summarized as top ten keyphrases.}
    \begin{tabular}{l|l}
    Keyphrase & Score \\ \hline
       presence  &   0.02041868080426608\\
       molecular weights  & 0.01313742352650019 \\
       glutamine synthetase & 0.01081927396059080\\
       growth hormone & 0.01081481738381907 \\
       arterial blood & 0.00973761662559790 \\
       investigated & 0.00926714499542069 \\
       rate constant & 0.00904369510973679 \\
       blood flow & 0.00899499866920862 \\
       molecular weight & 0.00865807865159297 \\
       sodium dodecyl & 0.00865611530561878
    \end{tabular}
    \label{tab:top-medal}
\end{table}
The top keyphrases correspond to rather general biological terms, which are some of the main topics related to the considered documents. The results were obtained by maintaining the merge\_threshold hyperparameter set to one -- single term keyphrases can be obtained if this threshold is lowered. For example, if set to 0.5, the top three keyphrases are `activity', `concentration' and `enzyme'.
\section{Discussion and conclusions}
\label{sec:discussion}
In this paper we presented an approach to unsupervised keyphrase detection, aimed specifically at pushing the limits of computation time and retrieval performance.
The main contributions of this paper are an algorithm for keyphrase detection that performs substantially (significantly) faster than current state-of-the-art methods, while maintaining the retrieval performance. The algorithmic novelties introduced touch upon the transformation of token sequences into graphs, and re-address the question of meta vertices by constructing them at the sequence level, which is substantially faster. Further, by exploiting personalized PageRank, global token information is incorporated into keyphrase ranking alongside token lengths. By conducting an extensive benchmark against established baselines, this paper presents an evaluation which incorporates both retrieval capabilities, but further details into computation time and duplication rates amongst the retrieved keyphrases.

Analysis of keyphrase detection algorithms with regards to multiple evaluation criteria is becoming of higher relevance, as many low-latency applications cannot afford expensive detection phase. To our knowledge, this paper is similarly one of the first to evaluate the performance based on critical difference diagrams, exactly assessing the significance of observed differences (in time and retrieval performance).

Further work includes exploration of lower-level implementations of top-performing approaches, alongside their parts that could be subject to parallelism. A potentially interesting endeavor would also include background knowledge (as graphs), possibly enabling detection of keywords beyond the ones found in a given document, while remaining unsupervised.

\section{Replicability}
The RaKUN 2.0 algorithm is available as a simple-to-use Python library available at \url{https://github.com/SkBlaz/rakun2}.

\section*{Acknowledgements}
The work was supported by the Slovenian Research Agency (ARRS) core research programme Knowledge Technologies (P2-0103), and projects Computer-assisted multilingual news discourse analysis with contextual embeddings (J6-2581) and Quantitative and qualitative analysis of the unregulated corporate financial reporting (J5-2554). The work was also supported by the Ministry of Culture of Republic of Slovenia through project Development of Slovene in Digital Environment (RSDO).
\bibliographystyle{splncs04}
\bibliography{mybibliography}

\begin{thebibliography}{10}
\providecommand{\url}[1]{\texttt{#1}}
\providecommand{\urlprefix}{URL }
\providecommand{\doi}[1]{https://doi.org/#1}

\bibitem{aronson2000nlm}
Aronson, A.R., Bodenreider, O., Chang, H.F., Humphrey, S.M., Mork, J.G.,
  Nelson, S.J., Rindflesch, T.C., Wilbur, W.J.: The nlm indexing initiative.
  In: Proceedings of the AMIA Symposium. p.~17. American Medical Informatics
  Association (2000)

\bibitem{augenstein2017semeval}
Augenstein, I., Das, M., Riedel, S., Vikraman, L., McCallum, A.: {S}em{E}val
  2017 task 10: {S}cience{IE} - extracting keyphrases and relations from
  scientific publications. In: Proceedings of the 11th International Workshop
  on Semantic Evaluation ({S}em{E}val-2017). pp. 546--555. Association for
  Computational Linguistics, Vancouver, Canada (2017).
  \doi{10.18653/v1/S17-2091}, \url{https://aclanthology.org/S17-2091}

\bibitem{overviewKeywordextractionGraphs}
Beliga, S., Meštrović, A., Martincic-Ipsic, S.: An overview of graph-based
  keyword extraction methods and approaches. Journal of Information and
  Organizational Sciences  \textbf{39},  1--20 (07 2015)

\bibitem{bird2009natural}
Bird, S., Klein, E., Loper, E.: Natural language processing with Python:
  analyzing text with the natural language toolkit. " O'Reilly Media, Inc."
  (2009)

\bibitem{boudin:2016:COLINGDEMO}
Boudin, F.: pke: an open source python-based keyphrase extraction toolkit. In:
  Proceedings of COLING 2016, the 26th International Conference on
  Computational Linguistics: System Demonstrations. pp. 69--73. Osaka, Japan
  (December 2016), \url{http://aclweb.org/anthology/C16-2015}

\bibitem{boudin2018unsupervised}
Boudin, F.: Unsupervised keyphrase extraction with multipartite graphs. In:
  Proceedings of the 2018 Conference of the North {A}merican Chapter of the
  Association for Computational Linguistics: Human Language Technologies,
  Volume 2 (Short Papers). pp. 667--672. Association for Computational
  Linguistics, New Orleans, Louisiana (2018). \doi{10.18653/v1/N18-2105},
  \url{https://aclanthology.org/N18-2105}

\bibitem{bougouin-etal-2013-topicrank}
Bougouin, A., Boudin, F., Daille, B.: {T}opic{R}ank: Graph-based topic ranking
  for keyphrase extraction. In: Proceedings of the Sixth International Joint
  Conference on Natural Language Processing. pp. 543--551. Asian Federation of
  Natural Language Processing, Nagoya, Japan (2013),
  \url{https://aclanthology.org/I13-1062}

\bibitem{CAMPOS2020257}
Campos, R., Mangaravite, V., Pasquali, A., Jorge, A., Nunes, C., Jatowt, A.:
  Yake! keyword extraction from single documents using multiple local features.
  Information Sciences  \textbf{509},  257--289 (2020).
  \doi{https://doi.org/10.1016/j.ins.2019.09.013},
  \url{https://www.sciencedirect.com/science/article/pii/S0020025519308588}

\bibitem{JMLR:v7:demsar06a}
Dem{\v{s}}ar, J.: Statistical comparisons of classifiers over multiple data
  sets. Journal of Machine Learning Research  \textbf{7}(1),  1--30 (2006),
  \url{http://jmlr.org/papers/v7/demsar06a.html}

\bibitem{ding-luo-2021-attentionrank}
Ding, H., Luo, X.: {A}ttention{R}ank: Unsupervised keyphrase extraction using
  self and cross attentions. In: Proceedings of the 2021 Conference on
  Empirical Methods in Natural Language Processing. pp. 1919--1928. Association
  for Computational Linguistics, Online and Punta Cana, Dominican Republic (Nov
  2021). \doi{10.18653/v1/2021.emnlp-main.146},
  \url{https://aclanthology.org/2021.emnlp-main.146}

\bibitem{gollapalli2014extracting}
Gollapalli, S.D., Caragea, C.: Extracting keyphrases from research papers using
  citation networks. In: Brodley, C.E., Stone, P. (eds.) Proceedings of the
  Twenty-Eighth {AAAI} Conference on Artificial Intelligence, July 27 -31,
  2014, Qu{\'{e}}bec City, Qu{\'{e}}bec, Canada. pp. 1629--1635. {AAAI} Press
  (2014), \url{http://www.aaai.org/ocs/index.php/AAAI/AAAI14/paper/view/8662}

\bibitem{grootendorst2020keybert}
Grootendorst, M.: Keybert: Minimal keyword extraction with bert. (2020).
  \doi{10.5281/zenodo.4461265}, \url{https://doi.org/10.5281/zenodo.4461265}

\bibitem{hasan2014automatic}
Hasan, K.S., Ng, V.: Automatic keyphrase extraction: A survey of the state of
  the art. In: Proceedings of the 52nd Annual Meeting of the Association for
  Computational Linguistics (Volume 1: Long Papers). pp. 1262--1273.
  Association for Computational Linguistics, Baltimore, Maryland (2014).
  \doi{10.3115/v1/P14-1119}, \url{https://aclanthology.org/P14-1119}

\bibitem{10.3115/1119355.1119383}
Hulth, A.: Improved automatic keyword extraction given more linguistic
  knowledge. In: Proceedings of the 2003 Conference on Empirical Methods in
  Natural Language Processing. pp. 216--223 (2003),
  \url{https://aclanthology.org/W03-1028}

\bibitem{10.5555/1859664.1859668}
Kim, S.N., Medelyan, O., Kan, M.Y., Baldwin, T.: {S}em{E}val-2010 task 5 :
  Automatic keyphrase extraction from scientific articles. In: Proceedings of
  the 5th International Workshop on Semantic Evaluation. pp. 21--26.
  Association for Computational Linguistics, Uppsala, Sweden (2010),
  \url{https://aclanthology.org/S10-1004}

\bibitem{krapivin2009large}
Krapivin, M., Autaeu, A., Marchese, M.: Large dataset for keyphrases extraction
   (2009)

\bibitem{kumar2022comprehensive}
Kumar, T., Mahrishi, M., Meena, G.: A comprehensive review of recent automatic
  speech summarization and keyword identification techniques. Artificial
  Intelligence in Industrial Applications pp. 111--126 (2022)

\bibitem{marujo2013keyphrase}
Marujo, L., Viveiros, M., da~Silva~Neto, J.P.: Keyphrase cloud generation of
  broadcast news (2013)

\bibitem{medelyan2009human}
Medelyan, O.: Human-competitive automatic topic indexing. Ph.D. thesis, The
  University of Waikato (2009)

\bibitem{10.5555/1699648.1699678}
Medelyan, O., Frank, E., Witten, I.H.: Human-competitive tagging using
  automatic keyphrase extraction. In: Proceedings of the 2009 Conference on
  Empirical Methods in Natural Language Processing. pp. 1318--1327. Association
  for Computational Linguistics, Singapore (2009),
  \url{https://aclanthology.org/D09-1137}

\bibitem{https://doi.org/10.1002/asi.20790}
Medelyan, O., Witten, I.H.: Domain-independent automatic keyphrase indexing
  with small training sets. ArXiv preprint  \textbf{abs/10.1002} (2010),
  \url{https://arxiv.org/abs/10.1002}

\bibitem{medelyan2008topic}
Medelyan, O., Witten, I.H., Milne, D.: Topic indexing with wikipedia. In:
  Proceedings of the AAAI WikiAI workshop. vol.~1, pp. 19--24 (2008)

\bibitem{mihalcea-tarau-2004-textrank}
Mihalcea, R., Tarau, P.: {T}ext{R}ank: Bringing order into text. In:
  Proceedings of the 2004 Conference on Empirical Methods in Natural Language
  Processing. pp. 404--411. Association for Computational Linguistics,
  Barcelona, Spain (2004), \url{https://aclanthology.org/W04-3252}

\bibitem{10.1007/978-3-540-77094-7_41}
Nguyen, T.D., Kan, M.Y.: Keyphrase extraction in scientific publications. In:
  Goh, D.H.L., Cao, T.H., S{\o}lvberg, I.T., Rasmussen, E. (eds.) Asian Digital
  Libraries. Looking Back 10 Years and Forging New Frontiers. pp. 317--326.
  Springer Berlin Heidelberg, Berlin, Heidelberg (2007)

\bibitem{ilprints422}
Page, L., Brin, S., Motwani, R., Winograd, T.: The pagerank citation ranking:
  Bringing order to the web. Technical Report 1999-66, Stanford InfoLab (1999),
  \url{http://ilpubs.stanford.edu:8090/422/}, previous number =
  SIDL-WP-1999-0120

\bibitem{papagiannopoulou2020review}
Papagiannopoulou, E., Tsoumakas, G.: A review of keyphrase extraction. Wiley
  Interdisciplinary Reviews: Data Mining and Knowledge Discovery
  \textbf{10}(2),  e1339 (2020)

\bibitem{schutz2008keyphrase}
Schutz, A.T., et~al.: Keyphrase extraction from single documents in the open
  domain exploiting linguistic and statistical methods. M. App. Sc Thesis
  (2008)

\bibitem{10.1007/978-3-030-31372-2_26}
{\v{S}}krlj, B., Repar, A., Pollak, S.: Rakun: Rank-based keyword extraction
  via unsupervised learning and meta vertex aggregation. In: Mart{\'i}n-Vide,
  C., Purver, M., Pollak, S. (eds.) Statistical Language and Speech Processing.
  pp. 311--323. Springer International Publishing, Cham (2019)

\bibitem{wan-xiao-2008-collabrank}
Wan, X., Xiao, J.: {C}ollab{R}ank: Towards a collaborative approach to
  single-document keyphrase extraction. In: Proceedings of the 22nd
  International Conference on Computational Linguistics (Coling 2008). pp.
  969--976. Coling 2008 Organizing Committee, Manchester, UK (2008),
  \url{https://aclanthology.org/C08-1122}

\bibitem{wen-etal-2020-medal}
Wen, Z., Lu, X.H., Reddy, S.: {M}e{DAL}: Medical abbreviation disambiguation
  dataset for natural language understanding pretraining. In: Proceedings of
  the 3rd Clinical Natural Language Processing Workshop. pp. 130--135.
  Association for Computational Linguistics, Online (2020).
  \doi{10.18653/v1/2020.clinicalnlp-1.15},
  \url{https://aclanthology.org/2020.clinicalnlp-1.15}

\end{thebibliography}

\end{document}